\documentstyle[12pt]{article}
\begin{document}
\setlength{\unitlength}{1mm}

{\hfill   November 1996 }

{\hfill    WATPHYS-TH-96/19}

{\hfill    hep-th/9612061} \vspace*{2cm} \\
\begin{center}
{\Large\bf Non-minimal coupling and quantum entropy} 
\end{center}
\begin{center}
{\Large\bf of black hole
}
\end{center}
\begin{center}
Sergey N.~Solodukhin\footnote{e-mail: sergey@avatar.uwaterloo.ca}
\end{center}
\begin{center}
{\it Department of Physics, University of Waterloo, Waterloo, Ontario N2L 3G1, 
Canada}  
\end{center}
\vspace*{1cm}
\begin{abstract}
Formulating the statistical mechanics for a  scalar field
with non-minimal $\xi R \phi^2$ coupling
in a black hole background we propose modification of the original 
't Hooft ``brick wall'' prescription. Instead of the Dirichlet condition
we suggest some scattering ansatz for the field functions at the horizon.
This modifies the energy spectrum of the system and allows one 
to obtain the statistical entropy dependent on the
non-minimal coupling. For $\xi<0$ 
the entropy renormalizes the classical Bekenstein-Hawking
entropy in the correct way and agrees with the result previously
obtained within the conical singularity approach.
For a positive $\xi$, however, the results differ.
\end{abstract}
\begin{center}
{\it PACS number(s): 04.70.Dy, 04.62.+v}
\end{center}
\vskip 1cm
\newpage
\baselineskip=.8cm
\section{Introduction}
\setcounter{equation}0
A common hope  during last years is that the study of quantum 
entropy of black hole can shed light on the problem of obtaining
a statistical meaning of the Bekenstein-Hawking entropy \cite{1},
\cite{2}. A number of different approaches was proposed (see reviews
in \cite{3}). 
According to 't Hooft \cite{4} (see also \cite{5}) 
the entropy arises from a thermal bath of 
quantum fields propagating just outside the horizon. This is a purely
statistical calculation treating the quantum thermal bath as a system 
characterized by some energy spectrum. Being in equilibrium at a temperature 
$T=\beta^{-1}$, the states of the system are distributed according to Gibbs.
An important feature of this calculation is that the density of  states
becomes infinite approaching the horizon. That is why 't Hooft introduced
a ``brick wall'', a fixed boundary staying at a small distance
$\epsilon$ from the horizon. Assuming that the quantum fields do not
propagate within this wall, 't Hooft imposed the Dirichlet
condition at the boundary. A reformulation of this model was suggested in
\cite{Myers} by using the Pauli-Villars regularization scheme.
Introducing a number of fictitious fields (regulators) of different statistics
and masses (dependent on UV regulator $\mu$), 
it was shown that this procedure not only
regularizes  the UV problem in the effective action but also automatically
implements a cutoff for the entropy calculation allowing one to remove
the ``brick wall''. Thus, the ``brick wall'' can be conveniently considered
\cite{SSS} as a fictitious boundary the role of which is just to make
the calculation simpler.

In an alternate approach the entropy arises from entanglement by means
of the density matrix obtained by tracing over modes of the quantum field
propagating inside the horizon \cite{6}, \cite{7}, \cite{8}, \cite{Kab1}.

The powerful method to calculate both the classical and quantum entropies
of a hole is to apply the Euclidean path integral approach
\cite{GH}. For an arbitrary field system
it entails  closing the Euclidean time coordinate with a period $\beta=T^{-1}$
where $T$ is the temperature of the system. This yields a periodicity
condition for the quantum fields in the path integral. In the black hole 
case for arbitrary $\beta$ this procedure leads to an effective
Euclidean manifold which has a conical singularity at the horizon that 
vanishes for a fixed value $\beta=\beta_H$. The entropy is
 calculated by differentiating the corresponding free energy with 
respect to $\beta$ and then setting $\beta=\beta_H$. This procedure was 
consistently carried out for a static black hole and resulted in obtaining 
the general UV-divergent structure of the entropy \cite{9}-\cite{FS}.

For a quantum matter minimally coupled to gravity these three methods to
calculate the entropy lead to the compatible results \cite{Myers}, \cite{SSS},
\cite{8}, \cite{Kab1}, \cite{10}, \cite{F}, \cite{FS} 
and predict the similar structure of the UV divergences of the entropy.
As was proposed by Susskind and Uglum \cite{10}, these divergences
are absorbed in the renormalization of  Newton's constant and, according to
\cite{FS}, in the renormalization of the higher curvature couplings in the 
effective action.
So far, however, there has not been  agreement between these approaches
for the quantum entropy due to a matter non-minimally coupled to
gravity. This problem was considered in \cite{Kabat}, \cite{S},
\cite{LW}, \cite{Kab}, \cite{BS}, \cite{FFZ}.

The peculiarity of the non-minimal coupling is easily seen in the
conical singularity approach \cite{S}. Consider the Euclidean path 
integral for a scalar field with the non-minimal operator
$-\Box_\xi \equiv -\Box +\xi R$. For an Euclidean manifold with a conical
singularity the scalar curvature behaves at the singularity as
$\delta$-function even if the regular part of the curvature vanishes.
This $\delta$-like potential modifies the spectrum of the operator
$-\Box_\xi$ and the resultant path integral. The similar phenomenon happens for
a manifold with boundaries \cite{BS}. In that case the operator $-\Box_\xi$
has a $\delta$-like potential concentrated on the boundary.
In result, the conical singularity method gives rise to the following
entropy due to the non-minimal quantum scalar matter \cite{S}:
\begin{eqnarray}
&&S_q={1\over 4} A_{\Sigma} (1-6 \xi )c_1 (\mu) \nonumber \\
&&+\{-{1\over 8\pi} ({1\over 6}-\xi)^2 \int_\Sigma R +
{1\over 45}{1\over 32\pi}\int_\Sigma (R_{\mu\nu}n^\mu_i n^\nu_i-
2R_{\mu\nu\alpha\beta}n^\mu_i n^\alpha_i n^\nu_j n^\beta_j) \} c_2(\mu )~~,
\label{1}
\end{eqnarray}
where $A_\Sigma$ is the area of the horizon $\Sigma$ and $\{n_i, \ i=1,2 \}$
is a pair of vectors orthogonal to $\Sigma$. The constants $c_1(\mu )$
and $c_2 (\mu )$  depend on the UV (energy) cutoff $\mu$ and for
large values of $\mu$ they
grow to leading order as $\mu^2$ and $\ln \mu^2$ respectively.

The divergences in (\ref{1}) take the correct form so that in 
combination with the bare entropy they are absorbed  in the 
renormalization of 
Newton's constant and the quadratic-curvature couplings in the effective 
action. In particular,  Newton's constant is renormalized as follows
\cite{BD}
\begin{equation}
{1\over G_R}={1\over G_B}+(1-6\xi )c_1 (\mu )~~.
\label{2}
\end {equation}
One can see that even in  flat space there still exists dependence on
$\xi$ in the quantum entropy (\ref{1}) and in the renormalization of 
Newton's constant (\ref{2}).

There are simple arguments showing that the quantum entropy in the
't Hooft approach has a different dependence on $\xi$.
Consider a black hole background satisfying the constraint
$R=0$. Then,  solutions of the field equation of motion
$(\Box-\xi R)\phi=0$ are the same as for $\xi=0$. Consequently,
the energy spectrum
of the system and the entropy are not effected by the non-minimal coupling.
The similar arguments are applicable for the entanglement entropy as well
to argue that this entropy also does not depend on $\xi$.
This  allowed some authors to make a conclusion
that ``entropy of  a  quantum matter non-minimally coupled to
gravity does not have a statistical meaning''.
An important circumstance missed in this sort of  arguments is that
a non-minimal matter possesses some non-trivial interaction
with the horizon (the importance of this interaction was argued in \cite{BS}).
The ``brick wall'' prescription is not appropriate to describe this
interaction.
The goal of this paper is to modify the 't Hooft approach in order to
get the correct $\xi$-dependence for the statistical entropy
and obtain the correspondence with the conical singularity
method.

An important point in our consideration is an idea that in the non-minimal
case we are obliged to impose  certain boundary condition 
the form of which directly follows from the form
of the non-minimal coupling. 
Therefore, our strategy is to replace the
Dirichlet boundary condition  appearing in the original 
't Hooft calculation by some $\xi$-dependent  condition.
A motivation for this, actually, comes from the Euclidean version
of the theory. Indeed, the presence in the operator $-\Box_\xi$
of the $\delta$-like potential concentrated on the boundary or 
at the conical singularity can be precisely re-formulated as imposing an
appropriate condition on the field functions on the boundary or at the
singularity respectively. In the Euclidean theory these
conditions are found to take a simple form \cite{SS}.
To make our condition in the Minkowskian space-time  more clear, note that
the standard Dirichlet condition for a wave equation
means that a wave is reflected by the boundary with the change of phase 
equal to $\pi$.
Analogously, what we propose is essentially an
ansatz for the field function near the horizon that describes 
scattering by the hole with some non-trivial change of phase. 
As a result of the scattering, in the spectrum of the system appear some
low-energy modes  the density of
which, being proportional to $\xi$,  grows as
$g^{-1/2}(\epsilon )$ approaching the horizon. Namely the contribution
of these modes to the statistical entropy is essential to obtain the
correct dependence on $\xi$ in agreement with (\ref{1}).
In principle, our calculation is quite similar to the standard statistical
consideration of the non-ideal gas \cite{LL}: 
the corrections to the thermodynamical quantities of the ideal 
gas are expressed via the two-particle scattering phase shift.
The only difference from our case is that in 
our model  a particle interacts with the horizon rather than 
with other particles.
It is important to note that we propose the scattering  not just for needs 
of the entropy calculation. We believe that it is an actual interaction 
between the non-minimal matter and the horizon that is dictated by the form
of the non-minimal coupling.

\bigskip

\section{WKB field function and boundary condition}
\setcounter{equation}0
A straightforward generalization of the approach \cite{4}, \cite{Myers}
is to consider the more general case of static black hole with
the spheri-symmetric metric written in the form:
\begin{equation}
ds^2=-g(r)dt^2+{1\over g(r)} dr^2+r^2 (d\theta^2+\sin^2\theta
d\varphi^2).
\label{3}
\end{equation}
Not assuming this metric to satisfy any concrete field equations, we suppose
that  $g(r)$ is a non-negative function having
a simple zero at $r=r_+$ where it behaves as $g(r)={4\pi\over \beta_H}
(r-r_+)+O((r-r_+)^2)$. This corresponds to the position of the outer event
horizon at $r=r_+$. The Hawking temperature calculated for the metric
(\ref{3}) is $T_H=\beta_H^{-1}$. Note that the scalar curvature $R$ for
 the metric (\ref{3}) is function of 
only radial coordinate $r$.

In this black hole background we consider a non-minimal scalar field
which satisfies the equation:
\begin{equation}
(\Box-m^2-\xi R)\phi=0~~.
\label{4}
\end{equation}
Expanding the wave function in spherical coordinates
$\phi=e^{\imath E t}Y_{lm}(\theta ,\varphi )f(r)$, we obtain the equation on
the radial function
\begin{equation}
E^2r^4f(r)+r^2g\partial_r (r^2g\partial_r f(r))-r^4g (m^2+\xi R+{l(l+1)\over r^2})
f(r)=0~~.
\label{5}
\end{equation}

The WKB approximation provides us with the following solution of the
Eq.(\ref{5}):
\begin{equation}
f(r)=\rho (r)~e^{\pm \imath \int^r {dr\over g}k(r)}~~, ~k(r)=\sqrt{E^2-(m^2+\xi R+
{l(l+1)\over r^2})g(r)}~~,
\label{6}
\end{equation}
which is valid in the region where $k^2(r)\geq 0$. It is clear that for 
large  mass $m$ this  is  space closely located to the horizon.
If $R$ takes a non-zero value at the horizon then this region can be 
approximated by some sort of the constant curvature
 space. However, the exact results
for such a space show \cite{BD} that the mass parameter in the solution
enters only in the combination $(m^2-{1\over 6}R)$.  The WKB approximation,
being well-defined for large values of $l$, does not give this result.
Correcting this, we arrive at the following wave function
\begin{eqnarray}
&&f(r)=\rho (r)~\left( e^{ \imath \int^r {dr\over g}k(r)}+Ae^{- \imath \int^r 
{dr\over g}k(r)}\right)
~~, \nonumber \\
&&k(r)=\sqrt{E^2-(M^2(r)+{l(l+1)\over r^2})g(r)}~~,
\label{7}
\end{eqnarray}
where we denote $M^2(r)=m^2-({1\over 6}-\xi )R$.
Since the quantity $M^2(r)$ is slowly varying near 
the horizon it can be considered  constant there.
The constant $A$ in (\ref{7})
is to be determined from the boundary condition.
The amplitude $\rho (r)$ is a slowly varying function of $r$. In what follows,
we neglect its derivatives with respect to $r$ and omit writing 
it in the formulas.

Consider a boundary $\Sigma_\epsilon$ staying at a small distance
$\epsilon$ from the horizon $\Sigma$. In the limit $\epsilon\rightarrow 0$ 
it approaches
the horizon $\Sigma_\epsilon \rightarrow \Sigma$. The parameter $\epsilon$
is assumed to be smaller\footnote{Explicitly, this means that $\mu (\epsilon
\beta_H )^{1/2}<<1$.} than any UV cutoff $\mu^{-1}$ appearing under
regularization of the quantum
field theory \cite{SSS}, and such quantities as $\mu^2 g(\epsilon )$
are considered to be negligible.
Therefore, the boundary $\Sigma_\epsilon$ plays an intermediate role
just to
simplify the consideration. The condition, in principle, can be
imposed directly on the horizon.

In order to arrive at an idea of the boundary condition to be imposed on
$\Sigma_\epsilon$ let us start with the following ``simple-minded''
condition:
\begin{equation}
(n^\mu\partial_\mu \phi-\xi \kappa \phi )|_{\Sigma_\epsilon}=0~~,
\label{8}
\end{equation}
where $n^\mu$ is vector normal to $\Sigma_\epsilon$ and $\kappa$ is the 
extrinsic curvature of $\Sigma_\epsilon$, $\kappa=\nabla_\mu n^\mu$.
Namely the condition of this type
 appears in the Euclidean version of theory \cite{SS}.
For the metric (\ref{3}) and sufficiently
 small $\epsilon$ the condition (\ref{8})
reads
\begin{equation}
\left( g(r)\partial_r f(r)-\xi^\star f(r)\right) |_{r=r_++\epsilon}=0~~,
\label{8'}
\end{equation}
where $\xi^\star=2\pi \beta^{-1}_H \xi$.
For the function (\ref{7}) it gives the expression for the constant $A$:
$A=e^{\imath \eta (k)}$. The phase $\eta (k)$ is defined as follows:
\begin{equation}
e^{\imath \eta (k)}\equiv {\imath k(\epsilon )-\xi^\star\over  
\imath k(\epsilon )+\xi^\star}~~,
\label{10}
\end{equation}
where $k(\epsilon )\equiv k(r=r_++\epsilon )$.

The condition we are looking for is a $\xi$-dependent modification of
the Eq.(\ref{8'}). Curiously enough, an appropriate condition
takes the form
\begin{equation}
\left( g(r)\partial_r -\xi^\star \right)^\nu f(r) |_{r=r_++\epsilon}=0~,
\nu=2\xi~~~,
\label{10'}
\end{equation}
where $\left( g(r)\partial_r -\xi^\star \right)^\nu$ is a pseudo-differential 
operator. It acts as follows
$$
\left( g(r)\partial_r -\xi^\star \right)^\nu 
e^{ \pm\imath \int^r {dr\over g}k(r)}
=\left( \pm \imath k-\xi^\star \right)^\nu e^{ \pm\imath \int^r {dr\over g}k(r)}~~,
$$
where we neglected the derivatives of $k(r)$. The condition (\ref{10'})
leads to the following constant
\begin{equation}
A=e^{\imath\nu\eta (k)}~e^{\imath \pi (\nu+1)}~~,
\label{9}
\end{equation}
where $\eta (k)= \arctan \left( 
{2k(\epsilon )\xi^\star\over k^2(\epsilon )-{\xi^
\star}^2} \right)$ is defined as in (\ref{10}). Remarkably, for $\nu=0$ ($\xi=0$) the Eq.(\ref{10'}) coincides
with the Dirichlet condition. 

With the condition (\ref{10'}) imposed, the WKB field function  takes the form
\begin{equation}
f(r)=\left( e^{\imath \int^r_{r_++\epsilon} {dr\over g}k}+e^{\imath \nu \eta (k)}e^{\imath \pi (\nu+1)}
e^{-\imath \int^r_{r_++\epsilon} {dr\over g}k} \right)~~.
\label{11}
\end{equation}
This is that ansatz which we propose for the field function near the horizon.
The function (\ref{11}) is valid in the region $r_++\epsilon \leq r
\leq r_E$, where $r_E$ is determined by the equation $k^2(r_E)=0$.
The precise meaning of the condition appears when we consider the 
field function at the horizon.
Introduce new coordinate $z=\int^r g^{-1}dr$. 
  Then, at the horizon ($z$ goes to $-\infty$) 
the function
(\ref{11})
becomes
\begin{eqnarray}
&&f(z)=e^{\imath E(z-z_0)}+e^{\imath \nu \eta (E)}e^{\imath \pi (\nu+1)}
 e^{-\imath E(z-z_0)}~~,
\nonumber \\
&&e^{\imath \nu \eta (E)}=\left({\imath E-\xi^\star\over \imath E+\xi^\star}
\right)^\nu~~,
\label{11'}
\end{eqnarray}
where $z_0$ is a constant,  and describes
scattering by the hole with change of phase $\nu \eta (E)+\pi (\nu+1)$.

\bigskip

\section{Density of states and entropy calculation}
\setcounter{equation}0 
In order to discretize the energy spectrum and simplify  counting
of  states
we impose also the Dirichlet condition  $\phi=0$  for
$r=r_E$. This gives  the quantization condition
\begin{equation}
2\int^{r_E}_{r_++\epsilon} {dr\over g(r)}k(r)=\nu \eta (k) +2\pi n +\pi\nu~~,
\label{s}
\end{equation}
where $n$ is an integer number. Inverting this, we get the  equation
\begin{eqnarray}
&&n=n(E,l)\equiv n_0(E,l)+n_1(E,l)~~, \nonumber \\ 
&&n_0(E,l)={1\over \pi}\int^{r_E}_{r_++\epsilon} {dr\over g}k(r)~~,\nonumber \\
&&n_1(E,l)=-{\nu \over 2\pi}\left( \arctan \left( 
{2k(\epsilon )\xi^\star\over k^2(\epsilon )-{\xi^
\star}^2} \right)+\pi \right) ~~,
\label{13}
\end{eqnarray}
solution of which gives the energy levels $\{ E_{n,l} \}$ at a fixed $l$. 

 Near the horizon the metric function can be approximated as $g(r)
\simeq {4\pi\over \beta_H} (r-r_+)$. In this approximation
 the integral in (\ref{13}) 
is calculated exactly. We find in result that
\begin{eqnarray}
&&n_0(E,l)\simeq {\beta_H \over 2\pi^2} \left( E~ arctanh \sqrt{1-{E_{min,l}^2
\over E^2}}-\sqrt{E^2-E_{min,l}^2} \right)~~, \nonumber\\
&&n_1(E,l)=-{\nu\over 2\pi} \left(\arctan \left( \xi^\star
{2\sqrt{E^2-E_{min,l}^2}\over E^2-E_{min,l}^2 -{\xi^
\star}^2} \right)+\pi \right)~~,
\label{*}
\end{eqnarray}
where $E^2_{min,l}\equiv (M^2+{l(l+1)\over r^2_+})g(\epsilon )$.
We see that $E=E_{min,l}$ is the minimal possible energy at a fixed $l$.
For $E\simeq E_{min,l}$ the behavior of the function $n(E,l)$ (\ref{13}) is
mainly determined by the component $n_1(E,l)$ while for large enough $E$ the 
component $n_0(E,l)$ becomes the leading one. For $\nu=2\xi$ the  
$n(E,l)$ is monotonically increasing (both for $\xi>0$ and $\xi<0$) 
function\footnote{ In general the 
behavior of the function $n_1(E,l)$ depends on
relative sign of $\nu$ and $\xi$. Namely, if $sign (\nu )=sign (\xi )$
the function $n_1(E,l)$ is increasing for $E \geq E_{min,l}$ while
in opposite case, $sign (\nu )=sign (\xi )$, it is decreasing. This affects
the behavior of the whole function $n(E,l)$  which is monotonically increasing
in one case ($sign (\nu )=sign (\xi )$) and has one minimum in other
case ($sign (\nu )=-sign (\xi )$). Here we assume that $\nu=2\xi$
and discuss the possible choice $\nu=-2|\xi |$ below in the Discussion
section.} taking  value $n(E,l)=\xi$ for $E=E_{min,l}$. 
The total number of modes with energy less than $E$ can be determined
by taking the quantity $(n(E,l)-n(E_{min,l}))$ and summing
over the degeneracy of the angular modes:
\begin{equation}
n(E)\equiv \int dl(2l+1)(n(E,l)-n(E_{min,l}))=n_0(E)+n_1(E)~~,
\label{n}
\end{equation}
where the sum over $l$ has been approximated by an integral, and this
integration runs over non-negative values of $l$ for which the
square roots $k(r)$ and $k(\epsilon )$ in the integrand are real.
After the integration we have
\begin{equation}
n_0(E)={2\over 3\pi}\int^{r_E}_{r_++\epsilon}dr {r^2\over g^2(r)}
(E^2-M^2g(r))^{3/2}
\label{14}
\end{equation}
and
\begin{eqnarray}
&&n_1(E)=-{\nu\over \pi}{r^2_+\over g(\epsilon)} \int_0^{\sqrt{E^2-M^2g(
\epsilon )}}dkk\arctan 
\left( {2k{\xi^\star} \over k^2-{\xi^\star}^2}\right) \nonumber \\
&&=-{\nu\over \pi}{r^2_+\over g(\epsilon)} \{ {\bar{k}^2(\epsilon )\over 2}
\arctan \left( {2\bar{k}{\xi^\star} \over \bar{k}^2-{\xi^\star}^2}\right)
+{\xi^\star \bar{k}(\epsilon ) }
-{{\xi^\star}^2} \arctan {\bar{k}(\epsilon )\over \xi^\star} \}~~,
\label{15}
\end{eqnarray}
where $\bar{k}(\epsilon )=\sqrt{ E^2-M^2 g(\epsilon )}$.

One can think about this system of particles as consisting of  two
components: the ordinary particles with the number $n_0(E)$ as in the
original 't Hooft
model and the scattering particles  with the number $n_1(E)$.
Remarkably, $n_1 (E)$ is proportional to the horizon area $A_+=4\pi r_+^2$.

 For $E\simeq E_{min}=Mg^{1/2}(\epsilon )$ the functions $n_0(E)$ and
$n_1(E)$ behaves as follows
\begin{eqnarray}
&&n_0(E)\simeq {4\over 15} {\beta_H \over \pi} r_+^2 {M^2\over E^4_{min}}
(E^2-E^2_{min})^{5/2} \nonumber \\
&&n_1(E) \simeq sign(\nu \xi ){\beta_H\over 2\pi^2}r^2_+{M^2\over E^2_{min}}
(E^2-E^2_{min})^{3/2}~~.
\label{**}
\end{eqnarray}
If $sign(\nu\xi )=+1$ the total number of modes  
$n(E)$ is the monotonically
increasing function  of $E$ and $E=E_{min}$ is actual minimum of energy. 

To determine the thermodynamics of this system, we consider the free energy
of a thermal ensemble of scalar particles with an inverse temperature
$\beta$
\begin{equation}
\beta F=\int_{E_{min}}^\infty dE {dn\over dE} \ln (1-e^{-\beta E})
= \beta F_0+\beta F_1~~,
\label{17}
\end{equation}
where we separate the contributions due to the 
ordinary modes and the scattering modes.

Applying Pauli-Villars regularization scheme for 
the   present four-dimensional scalar field theory, one introduces
five regulator fields $\{\phi_i,~i=1,...,5\}$ of different statistics
and masses $\{m_i,~i=1,...,5\}$ dependent on the UV cut-off
$\mu$ \cite{Myers}. Together with
the original scalar $\phi_0=\phi$ ($m_0=m$) 
these fields satisfy two constraints:
$\sum_{i=0}^5 \Delta_i=0$ and $\sum_{i=0}^5 \Delta_im_i^2=0$,
where $\Delta_i=+1$ for the commuting fields, and $\Delta_i=-1$
for the anticommuting fields. Additionally, we assume that all the fields
have the same non-minimal coupling $\xi_i=\xi,~i=0,...,5$.
Not deriving the exact expressions for $m_i$, we just quote here
the  asymptotes
\begin{eqnarray}
&&\sum_{i=0}^5 \Delta_im^2_i\ln m^2_i =\mu^2 b_1+m^2\ln{m^2\over \mu^2}
+m^2 b_2~~, \nonumber \\
&&\sum_{i=0}^5 \Delta_i\ln m^2_i =\ln{m^2\over \mu^2}~~,
\label{id}
\end{eqnarray}
where $b_1$ and $b_2$ are some constants, valid in the 
limit $\mu \rightarrow \infty$.
With contribution of each field added the free energy (\ref{17}) becomes
\begin{equation}
\beta\bar{F}=\sum_{i=0}^5 \Delta_i \beta F^i=\beta \bar{F}_{0}
+\beta \bar{F}_1~~.
\label{17'}
\end{equation}

For the scattering modes of a single field we have
\begin{eqnarray}
&&\beta F_1=\int_{E_{min}}^\infty dE {dn_1\over dE} \ln (1-e^{-\beta E})
\nonumber \\
&&=-{\nu \over \pi}{r^2_+\over g(\epsilon )}\int_{E_{min}}^\infty dE
E\arctan  \left( 
{2\sqrt{E^2-E_{min,l}^2}\xi^\star\over E^2-E_{min,l}^2 -{\xi^
\star}^2} \right)\ln (1-e^{-\beta E})~~.
\label{18}
\end{eqnarray}
A remarkable property of the expression (\ref{18})
is that for $\nu=2\xi$ it depends only on the absolute value of $\xi$. 
Integrating over $E$ in (\ref{18}) and focusing only on the 
divergent for small $\epsilon$  terms, we find
\begin{equation}
\beta F_1\simeq -|\xi | 
{r_+^2\over g(\epsilon )} (E_{min}^2\ln(\beta E_{min})-{1\over 2}E^2_{min})
-{|\xi | r_+^2\chi\over \beta^2 g(\epsilon )}~~,
\label{19}
\end{equation}
where $\chi$ is an independent on $M$ constant.
Summing contributions of each field $\phi_i$, we obtain
\begin{equation}
\beta \bar{F}_1\simeq -{|\xi |\over 2}
r^2_+\sum_{i=0}^5 \Delta_i M^2_i \ln M^2_i~~,
\label{19'}
\end{equation}
where $M^2_i=m^2_i-({1\over 6}-\xi )R$.

Our calculation of a part of the free energy due to the ordinary modes
essentially repeats for more general static metric (\ref{3})
the calculation presented in \cite{Myers}. In the expression
for the free energy of a single field we can integrate by parts and get
\begin{equation}
\beta F_0=-\beta \int_{E_{min}}^\infty {n_0(E)\over e^{\beta E}-1}dE~~.
\label{19''}
\end{equation}
Focusing on the divergences at the horizon in the expression for
the number of ordinary modes $n_0(E)$ (\ref{14}) we find
\begin{equation}
n_0(E)\simeq -{r^2_+\over \pi} \left( {2\over 3}({E\beta_H \over 4\pi})^3 
 C +  M^2  ({E\beta_H
\over 4\pi})\right) \ln {E^2\over E^2_{min}}+{2\over 3\pi}({\beta_H
\over 4\pi})r^2_+ E^3M^2 (E^{-2}_{min}-E^{-2})~~,
\label{21'}
\end{equation}
where $C=(R_{\mu\nu}n^\mu_i n^\nu_i-
2R_{\mu\nu\alpha\beta}n^\mu_i n^\alpha_i n^\nu_j n^\beta_j)|_\Sigma$
with notions as in (\ref{1}).

Substitution of this in (\ref{19''}), integration over $E$
and summation for all fields $\phi_i$ give us the total free energy due to
the ordinary modes:
\begin{equation}
\beta \bar{F}_0\simeq -{1\over 24} {\beta_H\over \beta}r^2_+ 
\sum_{i=0}^5 \Delta_iM^2_i\ln M^2_i 
-{1\over 45}{1\over 32}{\beta_H^3\over \beta^3} r^2_+ C 
\sum_{i=0}^5 \Delta_i\ln M^2_i~~.
\label{22}
\end{equation}

It is a manifestation of the mechanism discovered in \cite{Myers}
that the expressions (\ref{19'}) and (\ref{22}) are regular in the limit
$\epsilon \rightarrow 0$. There is  a precise cancelation of the 
divergences between the original scalar and regulator fields.
The resultant expressions, however, become dependent on the UV 
regulator $\mu$.

Altogether,  (\ref{19'}) and (\ref{22})   give the total free energy
(\ref{17'})
of the system. Calculation of the entropy $S=\beta^2\partial_\beta \bar{F}$
at the Hawking temperature $\beta^{-1}=\beta_H^{-1}$ gives
\begin{equation}
S={1\over 2}({1\over 6}+|\xi | )r^2_+\sum_{i=0}^5 \Delta_iM^2_i\ln M^2_i
+{1\over 45}{1\over 8} r^2_+ C \sum_{i=0}^5 \Delta_i\ln M^2_i~~.
\label{21}
\end{equation}
Using the definitions of $C$ and  $M^2_i=m^2_i-({1\over 6}-\xi )R$,
and assuming that the value of the scalar curvature $R$ at the horizon is 
much smaller
than each $m^2_i$, we arrive at the expression
\begin{eqnarray}
&&S={1\over 4} A_{\Sigma}~{1\over 12\pi} (1+6 |\xi | )
\sum_{i=0}^5 \Delta_im^2_i\ln m^2_i 
+\{-{1\over 8\pi} ({1\over 6}+|\xi |)
({1\over 6}-\xi) \int_\Sigma R \nonumber \\
&&+
{1\over 45}{1\over 32\pi}\int_\Sigma (R_{\mu\nu}n^\mu_i n^\nu_i-
2R_{\mu\nu\alpha\beta}n^\mu_i n^\alpha_i n^\nu_j n^\beta_j) \}
\sum_{i=0}^5 \Delta_i\ln m^2_i ~~.
\label{23}
\end{eqnarray}
This is our main result. In minimal case
($\xi=0$) and for the Reissner-Nordstrem
background Eq.(\ref{23}) coincides with the result of \cite{Myers}.

\bigskip

\section{Discussion of the result}
\setcounter{equation}0 
It should be noted that the expression (\ref{23}) is not exactly that
we were going to obtain for the entropy anticipating the complete
agreement with 
the result (\ref{1}) found within the conical singularity method. The 
dramatical difference of our result is that the statistical entropy
(\ref{23}) is not analytic with respect to the non-minimal coupling $\xi$. 
For negative $\xi$ it exactly reproduces the conical expression
(\ref{1}) while for positive $\xi$ the both quantities 
differ. This is well illustrated in the leading order. The conical
entropy (\ref{1}) then takes the form $S_q={1\over 4} A_{\Sigma}
(1-6\xi )c_1(\mu )$ and becomes negative for $\xi > {1\over 6}$.
This puzzling behavior has been discussed 
in \cite{LW}, \cite{S} and more recently in 
\cite{x}. In particular, it was noted that a statistical entropy defined as
$S=-Tr \rho \ln \rho$ for a density matrix $\rho$ can not  behave in this way
 being automatically positive. Therefore, may be it is not so 
surprising that our semi-classical computation of the statistical entropy
gives rise to the expression 
$S={1\over 4} A_{\Sigma}
(1+6|\xi | )c_1(\mu )$ which is always positive.

However, can we
modify in some way 
our calculation and reach the complete agreement with (\ref{1})?
At first sight, it seems possible if we assume that $\nu=-2|\xi |$ instead of
$\nu=2\xi$ . Then the free energy 
$\beta F_1$ (\ref{18}) looks becoming dependent on 
sign of $\xi$ and the entropy does too. 
However, in this case
the function $n(E,l)$ (\ref{13}), (\ref{*}) is not
monotonically increasing for $\xi >0$. Instead, it develops a minimum in a
point $\tilde{E}_{min,l}$ different than $E_{min,l}$. 
One observes the same behavior
for the function $n(E)$ (\ref{n}) 
having sense of the total number of modes with energy 
less than $E$. As is seen from (\ref{**}), for $sign (\nu\xi )=-1$ it decreases
at $E=E_{min}$, takes minimal value at some point $\tilde{E}_{min}$
and then monotonically increases. This behavior means that
for $\nu=-2|\xi |,~\xi>0$ we have to re-count the number of modes
which is no more given by the expression (\ref{n}). In result, we should get 
new function $n(E)$ which is monotonic and leads  for
$\xi>0$ to a positive entropy as well. 

The computation we present in this work sheds some light on the
origin of the $\xi$-dependent part of the entropy of the non-minimal scalar
field.
It is important to note that the main contribution to $\beta F_1$ 
 comes from modes with energy close to $E_{min}=
Mg^{1/2}(\epsilon )$. For this energy we may substitute
$\ln (1-e^{\beta E})\simeq \ln (\beta E)$ in (\ref{18})
and this is the divergence (if $E\rightarrow 0$)
that appears in (\ref{19}). It is interesting
that one may give an alternative simple 
calculation of $\beta F_1$ assuming existence
in the spectrum (\ref{s}) exactly\footnote{
Since $n$ is an integer it is better to suppose that $n=[\xi ]$ where $[\xi ]$
is integer part of $\xi$. For a big $\xi$ this is, however, not important.} 
 $n=|\xi |$ modes with the minimal energy 
$E_l=E_{min,l}\equiv (M^2+{l(l+1)\over r^2_+})^{(1/2)}g(\epsilon )$.
Indeed, summing then over  $l$ we find
\begin{eqnarray}
&&\beta F_1=\int dl(2l+1)|\xi | \ln (1-e^{\beta E_l})
={2|\xi |r^2_+\over g(\epsilon )} \int^\infty_{E_{min}=E_{l=0}}dEE
\ln (1-e^{\beta E})
\nonumber \\
&&\simeq -{|\xi |r^2_+\over g(\epsilon )}
E^2_{min} \ln (\beta E_{min})
\label{end}
\end{eqnarray}
that is exactly the expression (\ref{19}).
So, namely, these low-energy modes are responsible for the $\xi$-dependent
part of the entropy. The role of the boundary condition we propose
in the Section 2 is just to provide us with necessary number of these
modes.
Unfortunately, it is not clear that we have 
a well-posed problem with those boundary conditions and, for example, the Hamiltonian is obviously self-adjoint (though we deal with apparently 
positive energy spectrum). Therefore, it may happen that there  exists 
a more strict way
of getting the low-energy behavior we mentioned above within
a well-formulated self-adjoint extension of the scalar field
Hamiltonian.

The following conclusions are in order.

{\bf 1.} The quintessence in understanding the entropy
of the non-minimal scalar field is existence of the low-energy
modes number of which is governed by the non-minimal coupling
$\xi$.  In our approach these modes appear
due to the non-trivial scattering condition which we impose on the
scalar field at the horizon. 
In the limit $\epsilon\rightarrow 0$ (``brick wall''
removed) they presumably 
become zero modes of the Hamiltonian which are infinitely
degenerate due to the angle dependence. However, their status in
the well-defined Hamiltonian picture is not clear.

{\bf 2.} For negative $\xi$ the entropy (\ref{1}) may have a
statistical explanation within the procedure described in this paper.
However, we still lack this explanation for (\ref{1}) for
positive $\xi$. Moreover, the validity of the expression (\ref{1})
for $\xi>0$ is under question. More careful analysis \cite{SS}
shows that the conical method is not unambiguous and one may
be needed to re-consider   the way one obtains the expression
(\ref{1}) within this method.

{\bf 3.} For $\xi<0$ the divergences of the statistical entropy may be renormalized by the renormalization of  Newton's constant according to (\ref{2}).
However, for $\xi>0$ even after such a renormalization we still have
a divergence in the entropy behaving as $S_{div}=3|\xi |A_{\Sigma}
c_1(\mu )$. It is not clear in a renormalization of which quantity it may be absorbed.

{\bf 4.} Our result may be considered as a confirmation (for the non-minimally
coupled scalar field) of the
point that any physical quantum field (bosonic or
fermionic) must have a positive entropy. 
This, however, may have dramatical consequences for the 
understanding the origin of the black hole entropy in  theories of the induced gravity.
In this kind of theories \cite{Sakharov}
the Einstein gravity arises in the low-energy regime by averaging over
the constituent matter fields 
interacting with a background (classical) metric.
The set of these fields is specially arranged to make the induced
Newton's constant $G_{ind}$ UV finite. It is hoped \cite{Jacobson},
\cite{FFZ} that the finite
black hole entropy $S_{bh}={1\over 4G_{ind}}A_{\Sigma}$ can be 
induced in a similar way as a statistical entropy of the constituents. 
This hope is essentially
based on the possibility to make the black hole entropy finite
by making finite Newton's constant. However, this assumes that among the
constituents there should present 
fields carrying negative divergent entropy which compensates
the positive divergent entropy of other constituents. In a concrete
realization \cite{FFZ} 
of the induced gravity the role of those particles is played by the scalar
fields non-minimally coupled to gravity with positive $\xi$. However,
if each   constituent has a positive divergent entropy 
the total induced entropy is also positive and divergent even if the
induced Newton's constant is finite. Thus, we have an obvious problem
with inducing the Bekenstein-Hawking entropy. Possibly, the approached 
developed in \cite{FFZ} can be useful in resolving this problem.

One certainly needs understand better all these issues and we hope
to return to them in further publications.

\bigskip

\section*{Acknowledgments}
I wish to thank V.P.Frolov, D.Fursaev, R.C.Myers and A.Zelnikov
for helful comments.
I would like to thank the North Atlantic Treaty Organization (NATO)
for  support. 
In part this work is also supported by the 
Natural Sciences and Engineering Research
Council of Canada.

\end{document}